\documentclass[prd, preprint, 12pt]{revtex4-1}

\usepackage{amsmath}
\usepackage{amssymb}
\usepackage{setspace}
\usepackage{graphicx}
\usepackage{natbib}
\usepackage{float}

\begin{document}
 
 % ! TEX spellcheck
 %
%\ \vskip 1.0 in

\begin{center}
 {  {\bf General Relativity, Torsion, and Quantum Theory}}

%\smallskip

\vskip 0.1 in

{{\bf Tejinder  P. Singh}}

%\smallskip
{\it Tata Institute of Fundamental Research,}
{\it Homi Bhabha Road, Mumbai 400005, India}\\
{\tt tpsingh@tifr.res.in}\\
%\smallskip

%\vskip 0.5cm
\end{center}
\vskip 0.4 in

{ \it ``There is no doubt that quantum mechanics has seized hold of a beautiful element of truth and that it will be a touchstone for a future theoretical basis in that it must be deducible as a limiting case from that basis, just as electrostatics is deducible from the Maxwell equations of the electromagnetic field or as thermodynamics is deducible from statistical mechanics. I do not believe that quantum mechanics will be the starting point in the search for this basis, just as one cannot arrive at the foundations of mechanics from thermodynamics or statistical mechanics.''}

\rightline  {- Einstein (1936)}

\bigskip

\centerline{\bf ABSTRACT}

\noindent {We recall some of the obstacles which arise when one tries to reconcile the general theory of relativity with quantum theory. We consider the possibility that gravitation theories which include torsion, and not only curvature, provide better insight into a quantum theory of gravity. We speculate on how   
 the Dirac equation and Einstein gravity could be thought of as limiting cases of a gravitation theory which possesses torsion.}
 
\smallskip

\setstretch{1.1}

\noindent 

\smallskip
\smallskip

\setstretch{1.4}

\section{General Relativity and Quantum Theory}
Should one `quantize' the general theory of relativity [GTR]? There are various reasons to believe that applying the standard rules of quantum theory to GTR may not be the right way to arrive at a quantum theory of gravity. Here, we list some of these reasons:

(i) Should the gravitational field be quantized at all, or is it sufficient to have a semiclassical theory of gravity, which couples quantum matter fields to classical gravity? The answer to this question is not known, and it can be decisively settled only by experiment \cite{bahrami12015}. If we decide that quantum matter in general does not produce classical gravity, and there indeed is a quantum theory of gravity, then further issues arise:

(ii) The rules of quantum theory are written down after assuming that a background spacetime manifold and classical metric is given. Applying these quantum rules to the very metric whose existence was pre-assumed for writing the rules does not seem like a logical thing to do. Such an application may or may not lead to the correct theory.

(iii) Quantum theory as we understand it is incomplete. It depends on an external classical time, which is part of a classical spacetime geometry, which in turn is produced by classical matter fields. Classical fields are a limiting case of quantum fields. In this way quantum theory depends on its classical limit; this is unsatisfactory. There ought to exist an equivalent reformulation of quantum theory which does not refer to a classical time. Only such a reformulation can provide insights into a quantum theory of gravity
\cite{Singh:2012}.

(iv) Quantum theory suffers from the quantum measurement problem. When a quantum system interacts with a classical measuring apparatus, there is an apparent breakdown of the Schr\"{o}dinger equation and the principle of linear superposition. While the problem may be resolved by reformulating quantum theory as Bohmian mechanics; or by invoking environmental decoherence and the many worlds interpretation [provided the Born probability rule can be understood in this framework],  a deeper issue remains. When can the [vaguely defined] measuring apparatus be called classical, and why should quantum theory have to appeal to its own classical limit, in order to explain the outcomes of measurements? A more complete description, such as Continuous Spontaneous Localization, does away with the measuring apparatus, and in this new dynamics, linear superposition becomes an approximate principle, holding for astronomical times for microscopic systems, but for very small durations, when it comes to macroscopic objects. It is also possible that, since gravity becomes important for macroscopic objects, and since these objects are the ones for which there is an apparent breakdown of linear superposition,  collapse of the wave function is mediated by gravity. If this is the case, then we need to understand how gravity modifies quantum mechanics, before we can quantize gravity \cite{RMP:2012}.

A related aspect of quantum measurement is the apparent instantaneous nature of the collapse, whereby the quantum state changes instantaneously all over space, including over regions which are space-like separated from each other. This is confirmed by experiments. Although such effects cannot be used for signalling, they suggest some kind of acausal influence outside the light-cone. Our understanding of the relation between quantum theory, special relativity, and space-time structure, may have to be modified to have a better understanding of such an influence.

(v) The cosmological constant problem is a direct example of the conflict between general relativity and quantum field theory, and is a problem irrespective of whether or not cosmic acceleration is explained by a cosmological constant. The vacuum energy of quantum fields makes an enormous contribution to the cosmological constant, many many orders larger than the observed value, and we do not know why its gravitational effect is not observed. We need an improved understanding of the relation between GTR and quantum field theory, to sort this out.

(vi) The symmetry group of quantum field theory on Minkowski space-time is the Poincar\'e group [including Lorentz boosts and space-time translations], not the Lorentz group. Elementary particles are represented by irreducible representations of the Poincar\'e group, labelled by mass and spin. On the other hand, in GTR, which describes a curved spacetime, the local symmetry group on tangent spaces is the Lorentz group, not the Poincar\'e group. This is another stark example of conflict between GTR and quantum theory. 
Since mass is the source of gravity, the origin of particle masses 
in quantum field theory presumably has something to do with quantum gravity. Now if one takes the classical limit of quantum gravity, why should the symmetry group change from Poincar\'e  to Lorentz?  The local gauge theory of the Poincar\'e group is a theory of gravity which naturally includes both curvature, and a property known as  {\it torsion}. GTR is a special case of such a theory, in which torsion is set to zero by hand  \cite{hehl, Blagojevic, Hehl3}.  It seems reasonable that to arrive at quantum gravity, one should quantize, not GTR, but a gravitation theory with torsion. 

On the other hand, as we have seen above, there are difficulties related to a straightforward quantization of a classical theory of gravitation. What is more plausible is that both quantum theory and GTR need to be modified, before they can be used to make a quantum gravity, and GTR and quantum theory are themselves suitable limiting cases of quantum gravity. In the present article, we speculate in Section III as to how torsion could be of assistance in progressing towards such a goal. Before we do so, we very briefly review the vast body of known work on gravitation theories with torsion \cite{hehl, Blagojevic, Hehl3}.

\section{Gravitation Theories with Torsion}
If we use curvilinear coordinates on a spacetime manifold, the parallel transport of a vector $A^i$ along a given curve leads  to a change in its components, given by
\begin{equation}
dA^{i} = - \Gamma^{i}_{jk} A^{j}dx^{k}
\end{equation}
where $\Gamma^{i}_{jk}$ is known as the affine connection. There is no a priori reason for the affine connection to be symmetric in the index pair $(j,k)$, and its antisymmetric part
\begin{equation}
S_{{j}{k}}^{\ \ i} \equiv \frac{1}{2} (\Gamma^{i}_{jk} - \Gamma^{i}_{kj}) \equiv \Gamma^{i}_{[jk]}
\label{cartan}
\end{equation}
is known as Cartan's torsion tensor, and unlike the symmetric part, it transforms like a tensor. In GTR, the torsion tensor is {\it assumed} to be zero, and the connection is assumed to be symmetric  - of course there is no convincing fundamental motivation for this assumption. The best one can say is that setting  torsion to zero is consistent with all experiments to date, and leads to simpler field equations. It has been suggested that effects of torsion become significant in the vicinity of extreme situations such as ultra-high densities and gravitational singularities.  Of course it is well known that in GTR the symmetric part of the affine connection, known as the Christoffel symbols, represents the gravitational force, and vanishes in a locally inertial frame. Gravitation is produced by mass-energy, and according to torsion theories, torsion is produced by spin angular momentum. It is intriguing that the affine connection encapsulates the effect of mass as well as spin, in its symmetric and antisymmetric parts, respectively. 

The metric tensor $g_{ij}(x)$ is introduced so as to enable measurement of distances between points on the manifold, and if the length of a vector has to remain unchanged upon parallel transport, the covariant derivative of the metric must vanish. This relates the Christoffel symbols $\left\{\begin{array}{c}\sigma \\\mu\nu\end{array}\right\}$  to the metric, and the connection is now given by
\begin{equation}
\Gamma^{\sigma}_{\mu\nu} = \left\{\begin{array}{c}\sigma \\\mu\nu\end{array}\right\} - K_{\mu\nu}^{\ \ \sigma}\quad; \quad K_{\mu\nu}^{\ \ \sigma} \equiv - S_{\mu\nu}^{\ \ \sigma}
+ S_{\nu\ \mu}^{\ \sigma} - S^{\sigma}_{\ \mu\nu} = - K_{\mu\ \nu}^{\ \sigma}
\end{equation}
with  the Christoffel symbols defined as
\begin{equation}
\left\{\begin{array}{c}\sigma \\\mu\nu\end{array}\right\} = \frac{1}{2}g^{\sigma\lambda}\left(\partial_\mu g_{\nu\lambda} +\partial_\nu g_{\mu\lambda} - \partial_\lambda g_{\mu\nu}\right)
\end{equation}
and  $K_{\mu\nu}^{\ \ \sigma}$, known as the contortion tensor, depends on the metric and on torsion.
The spacetime is known as Riemann-Cartan spacetime, and if torsion vanishes we recover the better known Riemann spacetime. 

We can covariantly split the torsion tensor into a traceless part and a trace. The traceless part, known as the modified torsion tensor, and defined as
\begin{equation}
T^{\ \ \sigma}_{\mu\nu} = S^{\ \ \sigma}_{\mu\nu} + \delta^{\sigma}_{\mu} S^{\ \ \lambda}_{\nu\lambda} 
- \delta^{\sigma}_{\nu} S^{\ \ \lambda}_{\mu\lambda}
\end{equation}
plays a significant role in the field equations.

The commutator of two covariant derivatives of a vector introduces the Riemann tensor, but now the commutator depends on torsion as well: 
\begin{equation}
[\nabla_{\mu},\nabla_{\nu}]A^{\rho} = R_{\mu\nu\sigma}^{\ \ \ \ \rho} A^{\sigma} - 2S_{\mu\nu}^{\ \ \lambda}\nabla_{\lambda}A^{\rho}
\end{equation}
The Riemann tensor  depends on torsion as well as on the symmetric part of the connection. These two parts can be separated out and it can be shown that
\begin{equation}
R_{\mu\nu\sigma}^{\ \ \ \ \rho} = \tilde{R}_{\mu\nu\sigma}^{\ \ \ \ \rho} + \tilde{\nabla}_{\nu}K_{\mu\sigma}^{\ \ \rho} + K_{\mu\lambda}^{\  \rho}\; K_{\nu\sigma}^{\ \ \lambda} - K_{\nu\lambda}^{\  \rho}\; K_{\mu\sigma}^{\ \ \lambda} 
\end{equation}
where $\tilde{\nabla}$ defines the covariant derivative without torsion, and 
$\tilde{R}_{\mu\nu\sigma}^{\ \ \ \ \rho}$ is the Riemann tensor in Riemann space-time. The Riemann tensor is antisymmetric in the first two indices, and in the last two indices, but is no longer symmetric under the exchange of the first and second pairs; nor does it satisfy the cyclic identity. Thus it has thirty-six independent components, instead of the twenty independent components in Riemann spacetime. Of the thirty-six components, sixteen are in the Ricci tensor, which is no longer symmetric, and twenty are in the Weyl tensor. 

Torsion has geometric significance. If two infinitesimal vectors are parallely transported along each other, one does not get a closed parallelogram, with the non-closure being caused by torsion. Furthermore, there is an important analogy of torsion with the theory of defects in solids: when we compare geometry with defects, curvature is the analog of `disclinations' and torsion is the analog of `dislocations' \cite{Ruggiero2003}.

Next, one has to consider the field equations which determine curvature and torsion, thereby providing a generalisation of GTR. The simplest extension of GTR is obtained by adhering to the same Lagrangian density and action function as in GTR, except to replace the symmetric connection by the full connection which includes torsion. This minimal extension is known as the Einstein-Cartan-Sciama-Kibble (ECSK) theory \cite{hehl}. The action for the ECSK theory is 
 given by:
\begin{equation}
W=\int d^4x \sqrt{-g}\lbrace \mathcal{L}(\psi,\nabla\psi,g)+ \frac{R}{2k}\rbrace
\end{equation} 
with $k={8\pi G}/{c^4}$, $R$ the Ricci scalar, and $\psi$ represents matter fields.  We note that the matter Lagrangian density contains torsion through the covariant derivative.
The field equations are obtained by variation of the action with respect to $\psi$, $g_{\mu\nu}$, and $S_{{\mu}{\nu}}^{\ \ {\sigma}}$ or $K_{{\mu}{\nu}}^{\ \ {\sigma}}$:
\begin{equation}
\frac{\delta\left(\sqrt{-g}\mathcal{L}\right)}{\delta\psi}=0; \qquad
%\end{equation}
%\begin{equation}
\frac{\delta\left(\sqrt{-g}R\right)}{\delta g_{\mu\nu}}=-2k\frac{\delta\left(\sqrt{-g}\mathcal{L}\right)}{\delta g_{\mu\nu}}; \qquad
%\end{equation}
%\begin{equation}
\frac{\delta\left(\sqrt{-g}R\right)}{\delta K_{\mu\nu\sigma}}=-2k\frac{\delta\left(\sqrt{-g}\mathcal{L}\right)}{\delta K_{\mu\nu\sigma}}
\end{equation}
We can define the energy-momentum tensor $\sigma^{\mu\nu}$ in the usual way, and spin angular momentum $\tau^{\sigma\nu\mu}$ through the torsion:
\begin{equation}
\sigma^{\mu\nu}=\frac{2}{\sqrt{-g}}\frac{\delta\left(\sqrt{-g}\mathcal{L}\right)}{\delta g_{\mu\nu}}; \qquad
%\end{equation}
%\begin{equation}
\tau^{\sigma\nu\mu}=\frac{1}{\sqrt{-g}}\frac{\delta\left(\sqrt{-g}\mathcal{L}\right)}{\delta K_{\mu\nu\sigma}}
\end{equation}
The field equations are then given by
\begin{equation}
G^{\mu\nu}-\overset{\bigstar}{\nabla}_\lambda\left(T^{\mu\nu\lambda}-T^{\nu\lambda\mu}+T^{\lambda\mu\nu}\right)=k\sigma^{\mu\nu}
\end{equation}
and
\begin{equation}
T^{\sigma\nu\mu}=k\tau^{\sigma\nu\mu}
\end{equation}
where $\overset{\bigstar}{\nabla}\equiv\nabla_{\alpha}+2S_{\alpha\mu}^{\ \ \mu}$.

The first field equation can also be written as \begin{equation}
G^{\mu\nu}=k\Sigma^{\mu\nu}
\end{equation}
where $\Sigma^{\mu\nu}=\sigma^{\mu\nu}+\overset{\bigstar}{\nabla}_\lambda\left(\tau^{\mu\nu\lambda}-\tau^{\nu\lambda\mu}+\tau^{\lambda\mu\nu}\right)$. $\Sigma^{\mu\nu}$ can be shown to be identical to the canonical energy-momentum tensor.
Since the second equation connecting spin and torsion is algebraic, one can replace  torsion by spin and effectively cast out torsion from the formalism. Then one can split the Einstein tensor $G^{\mu\nu}$ into the Riemannian part $G^{\mu\nu}\left(\left\lbrace\right\rbrace\right)$ and its non-Riemannian part and replace the torsion terms in the non-Riemannian part in terms of spin, and arrive at the combined field equation given by
\begin{equation}
G^{\mu\nu}\left(\left\lbrace\right\rbrace\right)=k\sigma^{\mu\nu}
+k^2\left[4\tau^{\mu\lambda}_{\,\,\,\,\,\,[\rho}\tau^{\nu\rho}_{\,\,\,\,\,\,\lambda]}-2\tau^{\mu\lambda\rho}\tau^{\nu}_{\ \lambda\rho}+
\tau^{\lambda\rho\mu}\tau^{\ \ \nu}_{\lambda\mu}+\frac{1}{2}g^{\mu\nu}\left(4\tau^{\ \lambda}_{\gamma,\,[\rho}\tau^{\gamma\rho}_{\ \ \lambda]}+\tau^{\gamma\lambda\rho}\tau_{\gamma\lambda\rho}\right)\right]
\end{equation} 
This equation generalizes Einstein equations to incorporate the effect of torsion.  

The ECSK theory is an example of Poincar\'e gauge theories; the latter being theories which result from making the Poincar\'e group local. Such gauge theories {\it necessarily} include torsion, apart from curvature, and cover a wide class of theories (depending on the choice of action), and include ECSK and GTR as special cases. Given the fundamental significance of the Poincar\'e group, both in quantum theory and in classical mechanics, it seems natural to believe that the correct classical theory of gravitation includes torsion also, and reduces to GTR under situations where the effect of torsion is too small to be observable.  

On the other hand, if one takes the stance that fundamental spin is intrinsically a quantum feature, one might be led to ask if torsion has any role to play in classical theories. And if not, then how does one still retain the significance of the Poincar\'e group in classical theories of gravitation? In the next section we speculate on how this might be possible, and how gauge theories with torsion might contain within themselves the seeds of quantum theory.

\section{General Relativity, Torsion, and Quantum Theory} 
One of the possible suggestions for resolution of the quantum measurement problem is that collapse of the wave-function is caused by gravity \cite{RMP:2012,  Singh:2015}. This idea has been investigated seriously by Karolyhazy, Diosi, Penrose, their collaborators, and by others. Essentially, the idea is that if the space-time geometry has intrinsic uncertainty and fluctuations, these can cause decoherence of the wave-function of a quantum object propagating on such a background. This partially addresses the measurement problem, although it only explains decoherence, and not actual collapse. Typically, it is seen that gravity induced decoherence is more significant for more massive and larger (macroscopic) objects, thus inducing classical behaviour. And it is insignificant for microscopic objects, so that quantum theory holds for them with great precision, as one would expect. An important result in this context is due to Karolyhazy \cite{Karolyhazi:66,Karolyhazi:86}, which says that for a spherical object of radius $R$ and mass $m$, the critical length $a_c$ over which its wave-function remains coherent is given by
\begin{equation}
\frac{a_c}{R} \approx \left(\frac{L}{R_S}\right)^{2/3}\; \left(\frac{R_S}{R}\right)^{1/3}
\label{ratio}
\end{equation}
where $L$ is its Compton wavelength and $R_S$ its Schwarszschild radius. 

The physics behind this equation can be briefly summarized as follows. Karolyhazy shows that if one tries to measure a spacetime length $s$, then because of inherent quantum effects it possesses an uncertainty $\Delta s$ given by 
$(\Delta s)^3 \sim L_p^2 \; s$. This is representative of intrinsic fluctuations in the spacetime geometry, which are modelled by assuming that the metric is not that of classical Minkowski spacetime, but a stochastic metric whose mean is Minkowski, and variance such that it reproduces the length uncertainty $\Delta s$. When one considers the propagation of a quantum wave-function for an object according to the Schr\"{o}dinger equation, in the aforesaid stochastic background, it can be shown that the wave-function decoheres beyond the critical length scale $a_c$ given by (\ref{ratio}). Thus it is the stochastic fluctuations are responsible for decoherence.

An important lesson is learnt from here if we restrict to the case of a Schwarzschild black hole and set $R=R_S$ in (\ref{ratio}), giving 
\begin{equation}
\frac{a_c}{R} = \left(\frac{L}{R_S}\right)^{2/3}
\end{equation}
An object is characterized as macroscopic if the coherence length is smaller than its size: $a_c\ll R$, and the above formula then implies that for macroscopic objects $L\ll R_S, m\gg m_{pl}, R_S\gg l_{pl}$ and $L\ll l_{pl}$. On the other hand, an object is quantum if  the coherence length is larger than its size: $a_c\gg R$, and the above formula  implies that for microscopic objects $L\gg R_S, m\ll m_{pl}, R_S\ll l_{pl}$ and $L\gg l_{pl}$. The result suggests, as expected, that objects smaller than Planck mass cannot be thought of as classical black holes, and should instead be considered as particles which obey quantum theory. Thus one limit is classical GTR and classical mechanics, the other limit is quantum theory on flat spacetime, and for $m\sim m_{pl}$ we should have a new dynamics to which quantum theory and GTR are approximations. This expectation is further strengthened by noting that GTR by itself has no indicator that it holds only for $m\gg m_{pl}$, nor does quantum theory has anything in it to suggest that it holds only for $m\ll m_{pl}$. The dividing scale $m_{pl}$ can only come from the underlying intermediate theory. We now present some ideas towards arriving at such an intermediate theory, by looking for GTR and the Dirac equation as its two limiting cases \cite{Sharma2014, Sharma2014a}. [It turns out to be conceptually simpler to deal with the Dirac equation rather than the Schr\"{o}dinger equation, since both GTR and Dirac equation are linearly sourced by the mass]. It turns out that torsion plays an important role in the search for such an intermediate theory!

While Einstein equations and Dirac equations look very different from each other, they bear a striking structural similarity if expressed in the Newman-Penrose [NP] formalism, which uses the tetrad language to express the connection and the Riemann tensor in terms of the so-called spin coefficients, via the  Ricci identities \cite{Chandra}. In the NP formalism four null vectors are employed:  $\bf l, n, m$ and $\bf \overline{m}$ where ${\bf l}$ and ${\bf n}$ are real, and ${\bf m}$ and ${\bf \overline{m}}$ are complex conjugates of each other, and they are regarded as directional derivatives: 
\begin{equation}
D={\bf l}, \quad \Delta={\bf n}, \quad \delta = {\bf m}, \quad \delta^{*} = \overline{\bf m}
\end{equation}
The metric can be constructed from these null vectors of the tetrad. Ricci rotation coefficients (also known as spin coefficients) are the analog of the affine connection (for now assumed symmetric) and arise in the definition of the covariant derivatives of the four null vectors, just as the Christoffel symbols are defined in terms of derivatives of the metric. There are twelve complex spin coefficients,  denoted by standard symbols:
\begin{equation}
 \kappa, \sigma, \lambda, \nu, \rho, \mu, \tau, \pi, \epsilon, \gamma, \alpha, \beta
 \end{equation} 
The ten independent components of the Weyl tensor are denoted by five complex Weyl scalars
\begin{equation}
 \Psi_0, \Psi_1, \Psi_2, \Psi_3, \Psi_4
 \end{equation}
 while the ten components of the Ricci tensor are denoted by  four real scalars and three complex scalars
 \begin{equation}
 \Phi_{00}, \Phi_{22}, \Phi_{02}, \Phi_{20}, \Phi_{11}, \Phi_{01}, \Phi_{10}, \Lambda, \Phi_{12}, \Phi_{21}
 \label{Ricci}
 \end{equation}
 The Riemann tensor can be expressed in terms of Weyl scalars and Ricci scalars, and directional derivatives of the spin coefficients. This is done via eighteen complex equations, known as Ricci identities, and a typical Ricci identity takes the form
\begin{equation} 
D\rho - \delta^{*}\kappa = (\rho^2+\sigma\sigma^{*}) + \rho( \epsilon + \epsilon^{*})
 -\kappa^{*}\tau -\kappa (3\alpha +\beta^{*}-\pi) + \Phi_{00}
 \end{equation}
The Ricci tensor is determined from the Einstein equations, and the eighteen complex Ricci identities obey sixteen real constraints, known as eliminant conditions, because there are only twenty independent  components of the Riemann tensor.

Consider next the four Dirac equations for the four spinor components $F_1, F_2, G_1, G_2$:  these can also be written in the NP formalism, and a typical Dirac equation has the form \cite{Chandra}
\begin{equation}
(D+\epsilon - \rho) F_1 + (\delta^{*} + \pi - \alpha) F_2 = i\mu_{*} G_{1}
\label{D1}
\end{equation}
where $\mu_* = mc / \sqrt{2}\hbar$. Evidently the Dirac equations have a striking similarity with the Ricci identities, with both having a pair of derivatives of spin-coefficients / Dirac spinors.  Assuming that this similarity is not a coincidence, we can make contact between the Dirac equation and gravitational physics and recover the four Dirac equations  as special cases of the Ricci identities, provided we set eight of the spin coefficients to zero
\begin{equation}
\rho=\mu=\tau=\pi=\epsilon=\gamma=\alpha=\beta=0
\label{vanishspin}
\end{equation}
and make the following novel identification  between the four Dirac spinors and the remaining four non-zero spin-coefficients \cite{Sharma2014}
\begin{equation}
F_1 =\frac{i}{\sqrt{l_p}}\; \lambda, \quad F_2=-\frac{i}{\sqrt{l_p}}\;\sigma, \quad G_1=\frac{1}{\sqrt{l_p}}\;\kappa^{*}, \quad G_2=\frac{1}{\sqrt{l_p}}\;\nu^{*}
\label{match}
\end{equation}
The Dirac equations follow from the Ricci identities provided we assume relations between the Riemann tensor components and the Dirac mass \cite{Sharma2014}, a typical example of this kind being
\begin{equation}
\Phi_{20}  + \Phi_{01} = (\mu_* + \nu)\kappa^*
\label{EDNT1}
\end{equation}
Unfortunately however, it turns out that the sixteen constraints [the eliminant conditions] on the Ricci 
identities lead to undesirable constraints on the Dirac equation, and this particular idea for  the Einstein-Dirac correspondence does not work.

But there is a way to get rid of the troublesome eliminant conditions. We recall that there are thirty-six real (equivalently eighteen complex) Ricci identities. If we introduce torsion, there are thirty-six independent components to the Riemann tensor, and as a result there are no eliminant conditions imposed on the Ricci identities when torsion is present. Thus we assume, henceforth, that the affine connection is no longer symmetric. The Ricci tensor now has six additional components, denoted by  the three complex quantities ($\Phi_0, \Phi_1$ and $\Phi_2$).  The Weyl tensor has ten additional components, denoted by the real quantities ($\Theta_{00}, \Theta_{11}, \Theta_{22}, \chi$) and the complex quantities ($\Theta_{01}, \Theta_{02}, \Theta_{12}$). The spin coefficients now have an additional term due to torsion, and we use the following notation to represent the spin-coefficients \cite{Jogia}:
\begin{equation}
\kappa = \kappa^{\circ} +\kappa_1, \qquad \rho = \rho^{\circ} + \rho_1
\end{equation}
etc. with the part  $\kappa^{\circ}$ being the torsion-free part, and the part  $\kappa_1$ being due to torsion.

We next write down the eighteen complex Ricci identities, now with torsion included in the spin-coefficients. A typical example of the modified identities is \cite{Jogia}
\begin{eqnarray}
%\begin{align}
D\rho - \delta^{*}\kappa = &\ \rho (\rho + \epsilon + \epsilon^{*}) + \sigma \sigma^{*} - \tau \kappa^{*} - \kappa (3\alpha + \beta^{*} - \pi) + \Phi_{00} \nonumber  \\
&- \rho(\rho_1 - \epsilon_1 + \epsilon_{1}^{*} ) - \sigma \sigma_1^{*} + \tau \kappa_1^{*} + \kappa(\alpha_1 + \beta_{1}^{*} - \pi_1 ) + i\Theta_{00}
%\end{align}
\end{eqnarray}
Two limits are of interest. In one limit, the torsion part of the spin-coefficients is set to zero. In this limit the Ricci identities reduce to the ones discussed above, and if the source for the Ricci tensor is taken as the matter energy-momentum tensor, we recover GTR. In the opposite limit, the torsion free part of the spin-coefficients is set to zero, and  only the torsion part is retained. We now assume that eight of these torsion dominated spin-coefficients are zero, precisely as in (\ref{vanishspin}) above, and the remaining four non-zero spin-coefficients are assumed proportional to the Dirac spinors, as in (\ref{match}).  The Dirac equations then follow from the Ricci identities provided the Riemann tensor obeys the following conditions \cite{Sharma2014}:
\begin{equation}
\Phi_{20} + i\Theta_{20} + \Phi_{01} + i\Theta_{01} - \Psi_1 - \Phi_0  = \mu_*\kappa^* 
\label{dir1}
\end{equation}
\begin{equation}
\Phi_{21} + i\Theta_{21} + \Phi_2 - \Psi_3 + \Phi_{02} + i\Theta_{02}  = \mu_* \nu^* 
\label{dir2}
\end{equation}
\begin{equation}
i\Theta_{12} - \Phi_{12} + i\Theta_{00} - \Phi_{00} + \Phi_{2}^* -\Psi_{3}^* = -\mu_* \sigma 
\label{dir3}
\end{equation}
\begin{equation}
i\Theta_{10} - \Phi_{10} - \Phi_{0}^* - \Psi_{1}^* + i\Theta_{22} - \Phi_{22} = -\mu_* \lambda 
\label{dir4}
\end{equation}
Thus on the one hand we have the torsion-dominated limit, which are the Dirac equations, and on the other hand we have the gravity dominated limit, which are the Einstein equations. In the former case, gravity is absent (Minkowski space-time) and matter behaviour is quantum. In the latter case matter behaviour is classical, and gravity dominates over torsion. Thus we may conclude that there  must be a more general underlying theory in which the torsion-free part and the torsion part of the spin-connection are both present, and to which GTR and quantum theory are both approximations. Possibly,  GTR is the $m\gg m_{pl}$ approximation, and Dirac theory is the $m\ll m_{pl}$ approximation. Finding this underlying theory remains a major unsolved problem in this approach, at present.

Some insight into the structure and dynamics of the underlying theory maybe obtained by writing down the following heuristic action,  which is a sort of combination of the action for GTR and for the Dirac equation:
\begin{equation}
S = \frac{c^3}{G} \int d^4x \;\sqrt{-g} R + \hbar\int d^{4}x \; \sqrt{-g}\;\overline{\psi}(x) (i\gamma^{\mu}\partial_\mu\psi) 
- mc \int d^{4}x\; \sqrt{-g}\; \overline{\psi}{\psi}
\label{schematicaction}
\end{equation}
If we would like to obtain Einstein equations for a point particle out of this action, in the classical limit $\hbar\rightarrow 0$, then in the last term we can replace $\overline{\psi} \psi$ by a spatial three-delta function 
$\delta^{3}({\bf x})$ representing localization of the mass at a point.

Let us try to estimate the relative magnitudes of the integrands in these three terms of the action, by introducing characteristic lengths, and ignoring the  four volume element and the metric which are common to all the three terms. We assume there is a characteristic length $l$ associated with the system, and the curvature scalar maybe estimated as $R\sim 1/l^2$ and the first term is hence $T_1\sim c^3/Gl^2$, while the second term is $T_2\sim \hbar/l^4$ and the third term is $T_3 \sim mc/ l^3$. If $T_1$ dominates over $T_2$, then because of the resulting field equations we expect $T_1 \sim T_3$ and therefore
\begin{equation}
\frac{c^3}{G} \frac{1}{l^2} \sim \frac {mc}{l^3} \implies l \sim \frac{Gm}{c^2} \sim  R_S
\end{equation}
If $T_2$ dominates over $T_1$ and is order $T_3$ then
\begin{equation}
\frac{\hbar}{l^4}\sim \frac{mc}{l^3} \implies l \sim \frac{\hbar}{mc} \sim L
\end{equation}
We observe that $T_1\gg T_2$  suggests $R_S\gg L$ (the scale implied by $T_2$ should be negligible) and $T_1\ll T_2$ suggests $R_S \ll L$ (the scale implied by $T_1$ should be negligible). This suggests that when $T_1\sim T_2$ and $m\sim m_{pl}$ a dynamical description arising out  of a joint consideration of the Dirac action and the Einstein-Hilbert action for a particle of mass $m$ might be possible.

To make progress, we have to express the Dirac spinor in the term $T_2$ in terms of the torsion part of the connection, as done in (\ref{match}) above. At a more fundamental level, since the torsion is expressed by twelve spin-coefficients, the Dirac quantum state which is expressed by four spinor components, will have to be replaced by a new object which has twelve complex components - in principle this new object can be the torsion part of the connection.  The same would have to be done for the Dirac spinor in the third term $T_3$ in the action above. The curvature scalar in the first term in the action is also to be expressed in terms of the spin coefficients and their derivatives, and one has to investigate if the first and second terms of the action can be suitably combined into one term. A suitable variational principle has to be devised to arrive at the generalised field equations, to which GTR and the Dirac equation are approximations. Presumably this is a Poincar\'e gauge theory, because it has torsion as well as curvature. But it is not a classical theory: since we work with the complex spin coefficients rather than a real affine connection, there seems a possibility of making contact with quantum theory. Further investigations in this direction are currently in progress. It is intriguing that the symmetric and antisymmetric part of the connection carry information about gravity and the matter field, respectively.  

It is worth pointing out an important similarity of these ideas with the boundary-bulk gauge/gravity duality demonstrated in the context of the AdS/CFT correspondence. We have argued that classical GTR and flat space quantum theory should be limiting cases of an underlying theory, with both the limiting theories living in the same number of dimensions. In contrast, if we allow for different dimensions,  we learn from AdS/CFT that a flat space quantum theory living on the boundary is dual to a gravity theory in the bulk. In particular, if Eqn. (\ref{schematicaction})  is rewritten assuming  different dimensions for the gravity and Dirac sectors, it would appear to be the starting point of braneworld models: namely we have bulk
gravity in $d+1$ dimensions and quantum matter theory in $d$ dimensions. One
could consider a decoupling limit of such a model which in principle could
realize the premise of having GTR and quantum theory as limiting cases. This facet appears worthy of further investigation.

\medskip 

\noindent {\bf Acknowledgments:} I would like to thank Shreya Banerjee, Srimanta Banerjee, Sayantani Bera and Suman Ghosh for helpful discussions. This work is supported by a grant from the John Templeton Foundation (\# 39530).

\bigskip

\bigskip

\centerline{\bf REFERENCES}

%\bibliography{biblioqmts3}
\bibliography{biblioqmtstorsion}

\end{document}